\documentclass[fleqn,twoside]{article}
\usepackage{espcrc2}
\readRCS
$Id: espcrc2.tex,v 1.2 2004/02/24 11:22:11 spepping Exp $
\usepackage{graphicx}

\newcommand{\AmS}{{\protect\the\textfont2
  A\kern-.1667em\lower.5ex\hbox{M}\kern-.125emS}}

\title{The Cosmic Ray Mass Composition in the Energy Range $10^{15} - 10^{18}$
eV measured with the Tunka Array: Results and Perspectives
}

\author{V.V. Prosin\address[SINP]{Skobeltsyn
Institute of Nuclear Physics, Moscow State University, Russia}
\thanks{This work is supported by Russian Federation Ministry of Science and
Education (Contract No 02.518.11.7073), the Russian Foundation for Basic
Research (grants 07-02-00904, 05-02-04010, 06-02-16526) and the German Research
Foundation DFG(436 RUS 113/827/0-1). Correspondence to v-prosin@yandex.ru} for
the Tunka Collaboration:\\ 
\mbox{N.M. Budnev\address[APIISU]{Institute of Applied Physics, Irkutsk
State University, Irkutsk, Russia}}, 
\mbox{O.A. Chvalaiev\addressmark[APIISU]}, 
\mbox{O.A. Gress\addressmark[APIISU]},  
\mbox{N.N. Kalmykov\addressmark[SINP]}, 
\mbox{V.A. Kozhin\addressmark[SINP]},  
\mbox{E.E. Korosteleva\addressmark[SINP]},
\mbox{L.A. Kuzmichev\addressmark[SINP]}, 
\mbox{B.K. Lubsandorzhiev\address[INR]{Institute of Nuclear Research, Russian Academy
of Sciences, Moscow, Russia}},  
\mbox{R.R. Mirgazov\addressmark[APIISU]}, 
\mbox{G. Navarra\address{Dipartimento di Fisica Generale dell'Universita' and INFN,
Torino, Italy}},  
\mbox{M.I. Panasyuk\addressmark[SINP]}, 
\mbox{L.V. Pankov\addressmark[APIISU]}, 
\mbox{V.V. Prosin\addressmark[SINP]},
\mbox{V.S. Ptuskin\address{Institute of Terrestrial Magnetism, Ionosphere and
Radio Wave Propagation, Russian Academy of Sciences, Moscow, Russia}, 
\mbox{Yu.A. Semeney\addressmark[APIISU]}, 
\mbox{B.A. Shaibonov (junior)}\addressmark[INR]}, 
\mbox{A.A. Silaev\addressmark[SINP]}, 
\mbox{A.A. Silaev (junior)\addressmark[SINP]}, 
\mbox{A.V. Skurikhin\addressmark[SINP]}, 
\mbox{C. Spiering\address[DESY]{DESY, Zeuthen, Germany}}, 
\mbox{R. Wischnewski\addressmark[DESY]}, 
\mbox{I.V. Yashin\addressmark[SINP]}, 
\mbox{A.V. Zablotsky\addressmark[SINP]}, 
\mbox{A.V. Zagorodnikov\addressmark[APIISU]}
}
       
\begin{document}

\begin{abstract}

The final analysis of the Extensive Air Shower (EAS) maximum $(X_{max})$ depth
distribution derived from the 
data of Tunka-25 atmospheric Cherenkov light array in the energy range
$3\cdot 10^{15} - 3\cdot 10^{16}$ eV is presented.  
The perspectives of $X_{max}$ studies with the new Cherenkov light array
Tunka-133 of 1 km$^2$ area, extending 
the measurements up to $10^{18}$ eV, are discussed.
\vspace{1pc}
\end{abstract}

\maketitle

\section{INTRODUCTION}

The 
study of primary mass composition in the energy range $10^{15} -
10^{18}$ eV is of crucial importance for the understanding of the origin of
cosmic 
rays and of their propagation in the Galaxy. The change from light to heavier
composition with 
growing energy marks the energy limit of cosmic ray acceleration in galactic 
sources (SN remnants), and of the galactic containment. An opposite change from
heavy to light composition at 
higher energy would testify the transition from galactic to extragalactic
sources.  
Both changes are expected in the energy range of interest in the present
investigation. 
  
To study the mean composition we use the relation between the logarithm of
mass $\ln A$ and the depth $X_{max}$ of the extensive air shower (EAS)
maximum: 
$<X_{max}> \propto <\ln A>$ -- which is well-known from electromagnetic cascade
theory.
$X_{max}$ is derived for every event from the
steepness of the atmospheric Cherenkov light lateral distribution function
(LDF). 
We extend here the simplest method of performing such analysis (i.e.
of relating the average $<X_{max}>$ and $<\ln A>$ values) to the more correct
one based on the analysis of the whole $X_{max}$  distribution.

We summarize the results of the analysis of the Tunka-25 experiment, based
on 25 stations covering an  area of  0.1  km$^2$, and therefore tuned to operate
between $3\cdot 10^{15}$ and $3\cdot 10^{16}$ eV.
Both quoted methods show the beginning of composition change from light to heavy
at energies above $10^{16}$ eV.
To extend the study to higher energies, higher
statistics and consequently  larger sensitive areas and solid angles are needed.
Such an array, Tunka-133 (1 km$^2$ area) is now under construction close to its
predecessor Tunka-25. Last winter the first part of the new array operated for
about \mbox{270 h} during clean moonless nights. The first results and
perspectives of the new array are presented.

\section{TUNKA-25 FINAL RESULTS}

The detailed description of the Tunka-25 experiment and the procedure of
primary energy measurement and $X_{max}$ 
analysis from the LDF steepness is given in \cite{1}. The methods developed for
the analysis provided the relative accuracy of energy to be $\sim$15\% and the
error of $X_{max}$ less than 30 g/cm$^2$.
 
\begin{figure}[ht]
\includegraphics[width=75mm]{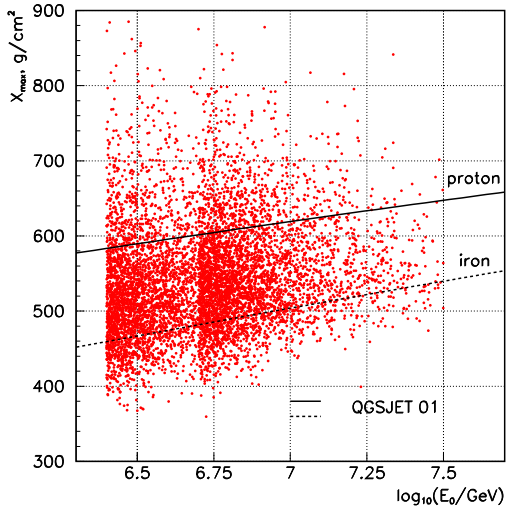} 
\caption{Depth of maximum, $X_{max}$, vs. primary energy $E_0$ (7632 points).}
\label{fig:1}
\end{figure}

\begin{figure}[ht]
\includegraphics[width=75mm]{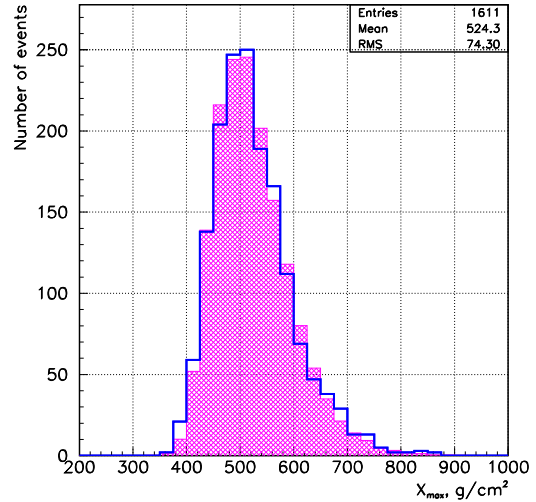} 
\caption{Depth of maximum, $X_{max}$, distribution for
$6.4 < log_{10}(E_0/GeV) < 6.5$. Line -- experiment, filled area -- simulation
for the complex mass composition containing $\sim$70\% of light (p+He) and
$\sim$30\%  of heavy (CNO+iron) nuclei}
\label{fig:2}
\end{figure}

The experimental plot of the depth of the shower maximum vs.
primary energy $E_0$ is shown in fig. 1. 
Data points are reported for energy $E_0 > 2.5\cdot 10^{15}$ eV at zenith
angles $\theta \geq 12^{\circ}$ 
and for $E_0 > 5\cdot 10^{15}$ eV at $\theta \geq 25^{\circ}$.
In fact, the analysis of
possible distortions of the distribution has shown that no
systematic errors are introduced in such energy-angular ranges.
The distributions inside narrow logarithmic
energy bins (0.1) have been analyzed. The analysis was done as follows. The
experimental $X_{max}$ distribution is compared with the simulated one. The
simulated distribution is constructed from 4 partial distributions for
different nuclei groups -- p, He, CNO and Fe. Partial distributions are
simulated with a "model of experiment" code assuming the QGSJET-01 model of
primary interaction. The code itself includes all the essential parameter
correlations and distributions extracted from CORSIKA and takes into account
all the apparatus errors and selection of events. A detailed description of
this code is given in \cite{1}. The weight of each group is selected for the
best 
fit of the experimental distribution. 

The result for one of the logarithmic
bins \mbox{($6.4 < log_{10}(E_0/GeV) < 6.5$)} is shown in fig. 2. We note that
QGSJET-01 provides the best fit of the left edge of the distribution when
compared with the other models. In principle the result of this procedure can
give the relative weight of each group of nuclei within the total composition.
But limited statistics and the relatively large width of the partial
distribution make it almost impossible to distinguish 2 components inside the
light (proton+helium) and the heavy (CNO+iron) groups.

\begin{figure}[ht]
\includegraphics[width=75mm]{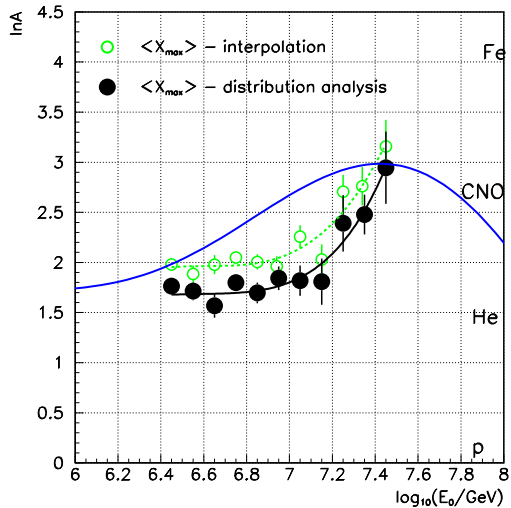} 
\caption{Mean logarithmic mass vs. primary energy. Dense curve is the
theory \cite{2}, dotted curves are the smoothing approximations of the
experimental points.}
\label{fig:3}
\end{figure}

A more stable 
solution can be obtained for the percentage of light (p+He) and
heavy (CNO+iron) nuclei in the primary composition, that provides a stable 
estimation of the mean logarithmic mass $<\ln A>$. The obtained experimental
dependence of $<\ln A>$ on primary energy $E_0$ is shown in fig. 3, together
with a  theoretical curve derived from \cite{2}. 

One sees that the old method of interpolation shifts $<\ln A>$ systematically by
about 0.25 towards heavier composition, compared to the more strict method of
the analysis of the full distribution. The mean value obtained with the second
method for the knee range of energies ($3\cdot 10^{15}$ eV) is close to that
obtained in the recent balloon experiments for energy about 10 TeV \cite{3}. The
rise of $<\ln A>$ for energies above $10^{16}$ eV is well visible.

\section{THE 1 km$^2$ TUNKA-133 ARRAY}

To study the mass composition behavior in the  energy range 
$10^{16}-10^{18}$ eV, the new array Tunka-133 is under construction \cite{4}.

\begin{figure}[ht]
\includegraphics[width=75mm]{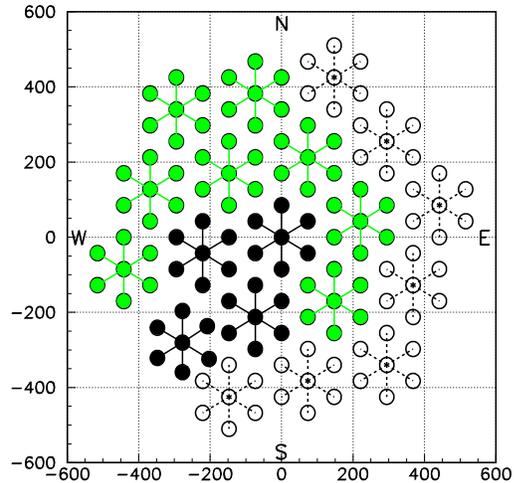} 
\caption{Plan of the Tunka-133 array. Black - 4 clusters operated in winter
2007-2008, grey -- 8 clusters ready for operation in winter 2008-2009, open
circles -- the last 7 clusters to be deployed in 2009.}
\label{fig:4}
\end{figure} 

The array will consist of 133 detectors grouped into 19 clusters each composed
of 7 detectors. The map of the array is shown in fig. 4. The new array
provides much more information than the previous one. Each detector signal is
digitized by an FADC with time step 5 ns. So the waveform of every pulse is
recorded, together with the preceding noise, as a total record of 5 $\mu s$
duration. 

The minimal pulse FWHM is about 20 ns and the dynamic range of
amplitude measurement about $10^4$. The latter is achieved by means of two
channels for each detector extracting the signals from the anode and an
intermediate dynode of the PMT with different additional amplification factors. 

Four clusters operated last winter between
November and April. Data have been recorded over 270 hours during clean
moonless nights. The average trigger rate was about 0.3 Hz, the number of the
registered events was about $3\cdot 10^5$.

\section{RECONSTRUCTION OF EAS PARAMETERS.}

\subsection{Structure of the data processing.}

The program of calibration and reconstruction of EAS parameters consists of
three main blocks.

1. The first block analyzes the primary data records for each Cherenkov light
detector and derives three main parameters of the pulse: front
delay at a level 0.25 of the maximum amplitude ($t_i$), pulse area ($Q_i$) and
full width on half-maximum FWHM$_i$.

2. The second block of codes combines the data of different clusters and provides
the relative time and amplitude calibration. Data from various
clusters are merged to one event, if the time 
difference for cluster triggers is less than 2 $\mu s$.

The time and amplitude calibration procedure is the same
described in  \cite{1}.

3. The third block of programs reconstructs the EAS core location, primary
energy and depth of  shower maximum. 
 
\begin{figure}[ht]
\includegraphics[width=75mm]{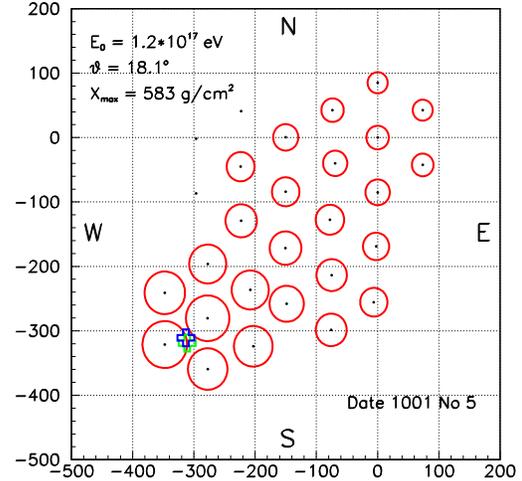} 
\caption{An example of an experimental event. The radii of the circles are
proportional to the logarithm of the Cherenkov light flux}
\label{fig:5}
\end{figure} 

\subsection {EAS core reconstruction with the density data $Q_i$} 

The first method of EAS core location reconstruction is based on the fitting of
the $Q_i$ data by the 
lateral distribution function (LDF) with varied parameters of steepness ($P$)
and light density at a core distance 175 m ($Q_{175}$). This function was first
suggested by the members of our collaboration in \cite{5}, and it is modified
here to include large distance measurements: $Q(R)=Q_{kn} f(R)$, 
$$
 f(R)=\left\{
\begin{array}{lc}
\!\exp\left(\frac{(R_{kn}-R)}{R_0}(1+\frac{3}{R+3})\right), R< R_{kn} \\
\!\left(\frac{R_{kn}}{R}\right)^{2.2},\hspace{15mm} 200\geq R\geq R_{kn} (1)\\
\!\left(\frac{R_{kn}}{200}\right)^{2.2}
\left((\frac{R}{200}+1)/{2}\right)^{-b}, R>200
\end{array} \right.
$$

Here $R$ is the core distance (in meters), $R_0$ is a parameter of the first
branch of LDF, $R_{kn}$ is the distance of the first change of LDF, $Q_{kn}$ is
the light flux at the distance $R_{kn}$. The second change occurs at the core
distance 200 m, $b$ is the parameter of the third branch. This branch is checked
till the distance 700 m with CORSIKA simulated events. 
These 4 variables are strictly connected with two main parameters of the LDF --
density at 175 m $Q_{175}$ and steepness $P$:
$$
\begin{array}{lc}
Q_{kn} = Q_{175} (R_{kn}/175)^{-2.2}\\
R_0 = \exp(6.79 - 0.564P),\hspace{4mm} (m)\\
R_{kn} = 207 - 24.5P,\hspace{14mm} (m)\\ 
b =\left\{
\begin{array}{lr}
4.84 - 1.23 ln(6.5-P), &  P < 6\\ 
3.43, &		P\geq 6
\end{array}
\right. 
\end{array}
$$


\begin{figure}[ht]
\includegraphics[width=75mm]{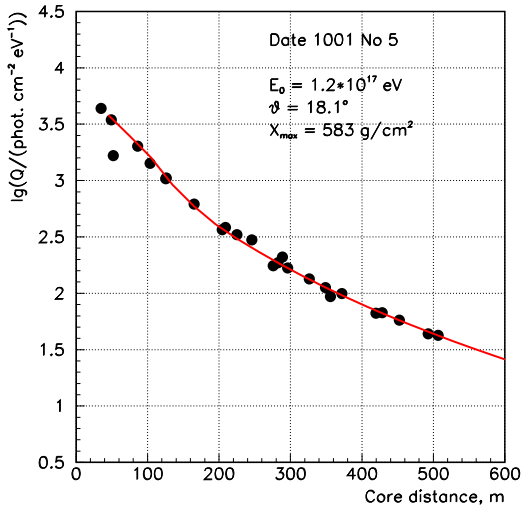} 
\caption{Lateral distribution resulting from fitting the measured light fluxes
(points) with the expression (1) (curve) for the event from fig. 5.}
\label{fig:6}
\end{figure}   

\begin{figure}[ht]
\includegraphics[width=75mm]{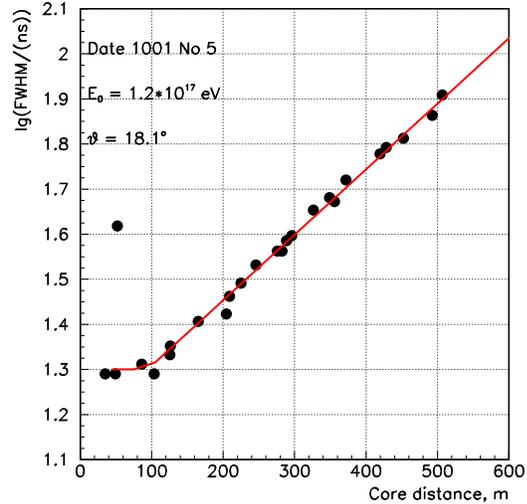} 
\caption{Dependence of Cherenkov light pulse width on the core distance for the
event from fig. 5. Points -- experiment, line -- approximation by expression
(2).} 
\label{fig:7}
\end{figure}

\subsection {EAS core reconstruction with the pulse widths FWHM$_i$}

In addition to the traditional method above described, a new method of EAS 
reconstruction using Cherenkov light pulse FWHM has been designed and included
into the analysis. To fit the experimental FWHM$_i$, the empirical
width-distance 
function (WDF) is used. It has a very simple analytic form (for FWHM $>$ 20 ns
and $R<500$ m): 
$$
\! \mathrm{FWHM}(R)=11\left(\frac{\mathrm{FWHM}(400)}{11}\right)^{\frac{R+100}{500}}ns
\eqno (2)
$$  

FWHM(400) is related to the depth of EAS maximum $X_{max}$, that will
therefore  
be reconstructed for each event by two
independent methods: from the LDF steepness $P$ and the parameter FWHM(400).

An example of a reconstructed shower is presented in fig. 5. The LDF and WDF
for this event are shown in fig. 6 and 7, respectively. 

\section{PERSPECTIVES OF THE PULSE WIDTH ANALYSIS}

 
The absence of FWHM random fluctuations and a simpler expression for the WDF
with respect to the LDF one seems to allow us applying the new method of EAS
core reconstruction not only inside, but also outside the array geometry, up to
a certain distance. 
 
A similar idea of reconstruction of EAS core distance using pulse width was
suggested many years ago by John Linsley \cite{6}. But the use of the idea for
charged particle detectors is problematic because of essential random
fluctuations of the signal form. Figure 7 shows that in case of Cherenkov
light random fluctuations do not play an essential role. 

We realize that before using such method of core reconstruction, a
detailed study of WDF up to core distances 1000 -- 1500 m has to be performed
not only with simulation but also experimentally. A positive result of
this study would let us expand the sensitive area of the array by 5 -- 10 times
compared with the geometrical area covered by the detectors for energies
above \mbox{$5\cdot 10^{17}$ eV}.
 
Such sensitive area increase will provide 20 -- 30 events with energies above
$10^{18}$ eV during one year (400 h during clear moonless nights) of observation
and ensure an overlapping of the Tunka-133 energy range with that of huge
installations such as Auger. 

\section{CONCLUSIONS} 

1. The analysis of $X_{max}$ distribution obtained with Tunka-25 array provides
the mean logarithmic mass of primary cosmic rays. For the energy range
$3\cdot 10^{15} - 10^{16}$ eV it is close to that obtained in 
balloon experiments for energies around $10^{13}$ eV and corresponds to a
"light" composition $\sim$70\% of p+He and $\sim$30\% of heavier nuclei.
For energies above $10^{16}$ eV a rapid growth of $<\ln A>$ is observed. 

2. The preliminary results obtained with the first stage of the new array
Tunka-133 show unique possibilities of the new apparatus recording pulse
waveform for each detector. This will provide more reliable measurement of
$X_{max}$ with two methods in each individual event. 

3. The completion of Tunka-133 with its projected area of 1 km$^2$, together
with the
development of the new method of shower core reconstruction from pulse
durations,
will provide the reliable evaluation of $<\ln A>$ up to $10^{18}$ eV.

\end{document}